\documentclass[aps,prc,twocolumn,showpacs,floatfix]{revtex4-1}
\usepackage{amsmath}
\usepackage{epsfig}
\usepackage{color}
\usepackage{endnotes}
\usepackage{natbib}
\usepackage[colorlinks,breaklinks]{hyperref}
\usepackage{nicefrac}
\usepackage{bm}
\usepackage{dcolumn}
\usepackage{graphics}
\usepackage{amsmath}
\usepackage{amssymb}
\usepackage{color}
\newcommand{\bra}{\langle}
\newcommand{\ket}{\rangle}

\newcommand{\be}{\begin{equation}}
\newcommand{\ee}{\end{equation}}
\newcommand{\bea}{\begin{eqnarray}}
\newcommand{\eea}{\end{eqnarray}}

\DeclareMathOperator*{\SumInt}{%
\mathchoice%
  {\ooalign{$\displaystyle\sum$\cr\hidewidth$\displaystyle\int$\hidewidth\cr}}
  {\ooalign{\raisebox{.14\height}{\scalebox{.7}{$\textstyle\sum$}}\cr\hidewidth$\textstyle\int$\hidewidth\cr}}   
  {\ooalign{\raisebox{.2\height}{\scalebox{.6}{$\scriptstyle\sum$}}\cr$\scriptstyle\int$\cr}}
  {\ooalign{\raisebox{.2\height}{\scalebox{.6}{$\scriptstyle\sum$}}\cr$\scriptstyle\int$\cr}}  
}

\begin{document}

\title{An efficient method for evaluating energy-dependent sum rules}

\author{N. Nevo Dinur}
\email{nir.nevo@mail.huji.ac.il}
\affiliation{Racah Institute of Physics, The Hebrew University, Jerusalem 91904, Israel}

\author{C. Ji}
\email{jichen@triumf.ca}
\affiliation{TRIUMF, 4004 Wesbrook Mall, Vancouver, BC V6T 2A3, Canada}

\author{S. Bacca}
\email{bacca@triumf.ca}
\affiliation{TRIUMF, 4004 Wesbrook Mall, Vancouver, BC V6T 2A3, Canada}
\affiliation{Department of Physics and Astronomy, University of Manitoba, Winnipeg, MB, R3T 2N2, Canada }

\author{N. Barnea}
\email{nir@phys.huji.ac.il}
\affiliation{Racah Institute of Physics, The Hebrew University, Jerusalem 91904, Israel}

\date{\today}

\begin{abstract}
Energy-dependent sum rules are useful tools in many fields of physics.
In nuclear physics, they typically 
involve an integration of the response function over the nuclear spectrum 
with a weight function 
composed of integer powers of the energy. 
More complicated weight functions are also encountered, 
e.g., in nuclear polarization corrections of atomic spectra.
Using the Lorentz integral transform method and the Lanczos algorithm, 
we derive a computationally efficient technique for evaluating such sum rules 
{that avoids the explicit calculation of both} 
the continuum states 
{and} 
the response function itself.
Our numerical results for electric dipole
sum rules of the $^4$He nucleus with 
various energy-dependent weights 
show rapid convergence with 
respect to the number of Lanczos steps.
This demonstrates the usefulness of the method in a variety of 
electroweak reactions. 
\end{abstract}

\maketitle

\section{Introduction}

Since the introduction of the Thomas-Reiche-Kuhn~\cite{TRK} and the Bethe~\cite{Bethe:1930ku} sum rules, quantum mechanical sum rules have been widely applied to  many fields of physics, including 
fundamental particles~\mbox{\cite{Weinberg1967,Shifman1998}}, 
atomic nuclei~\mbox{\cite{LSreport,OTreport,4HeSR}} 
and nuclear matter~\mbox{\cite{Polls1994,Mallik2010,Drukarev2012}}.
The definition of the term ``sum rule'' (SR) may vary in different physical problems. In the studies of atomic nuclei, energy-dependent sum rules are often referred to the spectral integration over a nuclear response function with an energy-dependent weight function. The response function is associated with transitions between the ground and excited states of a nucleus due to an external  probe. 
The SRs reflect important information about the structure and properties of nuclei. 
They are also key ingredients in calculating 
nuclear polarization effects on the spectrum of muonic atoms, 
which have recently drawn renewed interest~\mbox{\cite{Pachucki:2011xr,Antognini:2011zz,Ji2013a,Ji:2013vya,Friar:2013rha}}.

Since the response function is related to the reaction cross section
through kinematic factors, 
the corresponding SR can be evaluated through an integration over the appropriate cross section obtained from experiments.
However, when there are no available data,
or when the measurements are either limited in energy range or too scattered for a reliable estimation of 
the response function,
one must rely on theory for evaluating the SRs.
In many cases SRs can be directly obtained from ground-state expectation values of operators, 
and in some special cases they can even be evaluated in a model independent or quasi-independent way~\cite{Kirson1978,Li2012}.     
Best known examples are the Thomas-Reiche-Kuhn sum rule~\cite{TRK}
and the bremsstrahlung sum rule~\mbox{\cite{LevingerBethe,Brink,Foldy,DellafioreBrink}}.
They are associated with the first two spectral moments of the unretarded electric dipole response function, 
which corresponds to the dominant contribution in nuclear photoabsorption reactions.

In general, a sum rule may contain 
a weight function in a rather complex form
(see, e.g.,~\mbox{\cite{Pachucki:2011xr,Ji2013a,Friar:2013rha}}) and 
cannot simply be written as a ground-state expectation value.
In such cases, a theoretical evaluation of the SR requires a calculation of the response function, 
followed by an explicit integration over all the relevant excitation spectrum. 
An {\it ab-initio} solution of all the continuum states is a rather challenging task, often out of reach. 
Therefore, using indirect methods, such as the Lorentz integral transform (LIT)~\cite{EFROS94,REPORT07}, 
is presently the only viable way for calculating the response functions in an ab-initio approach (c.f.~\cite{Bacca2013} for a recent update), 
and consequently the SR.
Even then, obtaining accurate results from 
an explicit integration of the response function may be a rather demanding task, especially when the SR is
relatively sensitive to the high-energy part of the response function.

Here, we use the Lanczos algorithm~\cite{Lanczos} to 
derive a generalized numerical technique for evaluating SRs, 
dubbed `the Laczos sum rule (LSR) method',
which avoids the complications associated with explicitly calculating the continuum states or even
the response function. 
Similar LSR methods have been applied to study sum rules associated with Gamow-Teller and 
electromagnetic transitions (see, e.g.,~\cite{Haxton,Caurier:1994xg,Caurier:1998zw,4HeSR}), 
whose spectral weights only included integer powers of the energy.

In essence, in this paper we extend the sum rule method 
applied in Ref.~\cite{4HeSR} for evaluating the electric polarizability~\cite{Friar1975} 
to SRs with energy-dependent weight function of a more general form.

We define a general SR 
\begin{equation}\label{eq:SR}
I=\int_{0}^{\infty} d\omega\, S\left(\omega\right)g\left(\omega\right)\;,
\end{equation}
where $g\left(\omega\right)$ is the energy-dependent weight function, and $S(\omega)$
is the response function describing the excitation probabilities of
a nucleus due to some external probe. 
The response function is given by
\begin{equation}
\label{eq:S_omg}   
   S(\omega)=\SumInt_f |\bra f |\hat{O}| i \ket|^2
    \delta\left(E_{f}-E_{i}-\omega\right)\;,
\end{equation}
where
$\hat O$ is the transition operator 
(e.g., an electric dipole or quadrupole operator).
$|i\ket$ and $|f\ket$ are the initial and final eigenstates of the Hamiltonian $\mathcal{H}$ 
with eigenvalues $E_i$ and $E_f$, respectively. 
The $\SumInt$ symbol in Eq.~(\ref{eq:S_omg}) 
indicates a sum over the discrete part of the spectrum 
and an integral over the continuum states.
Throughout our discussion, the weight function $g\left(\omega\right)$ can have an arbitrary analytical form, 
restricted to only two conditions: 
$(i)$ the integrand \mbox{$S(\omega)g(\omega)$} in Eq.~(\ref{eq:SR}) is regular, 
and 
$(ii)$ the integral~\eqref{eq:SR} exists and is finite.

The lower bound of the integral \eqref{eq:SR} is the lowest excitation energy of the system, 
\mbox{$\omega_{th}\! \geq\!0$}. 
The upper bound is unlimited, however, for a given desired accuracy 
it can be regarded finite, since $S(\omega)$ vanishes for \mbox{$\omega \rightarrow \infty$}.

To calculate $|i\ket$ and $|f\ket$ one often expands them on 
a complete discrete set of localized basis states,  
effectively truncated at some finite basis size $M$, 
and then diagonalizes 
the Hamiltonian $\mathcal{H}$, 
obtaining a set of eigenstates $|\mu\ket$ and corresponding eigenvalues $E_\mu$. 
The states $|\mu\ket$  are not real eigenstates of the physical Hamiltonian,
but can be regarded as the discrete set of eigenstates that emerge when the 
system is confined in a finite volume.
For energies below the continuum threshold $E_\mu < E_{th}$,
the states $\{|\mu\ket\}$ are controlled approximations of the true bound states 
that improve by increasing the basis size $M$. 
For $E_\mu > E_{th}$,
these states are viewed as a discretized
approximation of the continuum that gradually fills in the continuum as the 
basis size is increased.
In terms of this discrete basis, the SR of Eq.~(\ref{eq:SR}) becomes
a sum over the transition probabilities from the
ground state to the discretized excited states, 
weighted by the function \mbox{$g(\omega)$}:
\begin{equation}
\label{I_sum}
  I_M 
   =\sum\limits_{\mu}^{M} |\bra \mu | \hat O | i\ket|^2 g(\omega_{\mu})\;,
\end{equation}
where $\omega_{\mu}=E_{\mu}-E_i$ is the excitation energy and the sum runs over all states but the ground state.
Here, the index $M$ denotes the fact that we are working with discrete states in a finite model space of size $M$.
Naively, $I_M$ {may be expected} to converge to the SR's physical value, $I$, 
as $M$ increases.
However, the verity and rate of the convergence 
\mbox{$I_M \rightarrow I$} 
require further investigation.

Additionally, in many calculations 
a complete diagonalization of the Hamiltonian is computationally impractical.
The Lanczos algorithm~\cite{Lanczos} can be introduced to 
handle the problem of large-model-space diagonalization.
Based on a recursive mapping of the full $M\times M$ Hamiltonian matrix into a tridiagonal matrix $\hat{H}_N$ 
whose dimension $N$ is much smaller than $M$,
the Lanczos algorithm effectively preserves information on 
the low-lying eigenstates and spectral moments~\cite{Haxton,Caurier:1994xg}.
Using the Lanczos algorithm, the SR in Eq.~\eqref{I_sum} becomes
\begin{equation}
\label{SR_lanc}
I_N = \bra i |\hat O^\dagger \hat O 
         |i\ket \sum_{\nu=0}^{N-1}| Q_{\nu 0}|^2 g(\omega_\nu) \;.
\end{equation}
Here the index $N$ denotes the number of Lanczos iterations, 
$Q$ is the unitary transformation matrix that diagonalizes $\hat{H}_N$,
and \mbox{$\omega_\nu \equiv E^{(N)}_\nu - E_i$}, where  $E^{(N)}_\nu$ is 
the $\nu$-th eigenvalue of $\hat{H}_N$.

Due to the advantage of the Lanczos algorithm~\mbox{\cite{Kaniel1966,Paige1971,Saad1980}}, 
$I_N$ converges rapidly to $I_M$, i.e., $\left|I_N - I_M \right|\rightarrow 0$ for 
a sufficiently large $N$, which turns out to be much smaller than the size of the discretized basis $M$. 
Therefore, when $I_M\rightarrow I$ is established, we obtain that $I_N \rightarrow I$, 
where $N$ can be kept relatively small. 
Thus, Eq.~(\ref{SR_lanc}), as the main point of this paper,
provides an efficient way to calculate SRs using the Lanczos algorithm, 
where the weight functions are 
more generalized than 
just integer powers of $\omega$ as in the traditional first moments of a distribution~\cite{Haxton}.

In the rest of the paper we 
{will present a theoretical derivation of} 
Eq.~(\ref{SR_lanc}). 
Using the properties of Lorentz integral transform (LIT), we will discuss 
the conditions for which
$I_M$, and therefore also $I_N$, converge to $I$. 
In particular, in 
Sec.~\ref{Derivation} we will prove that 
$(i)$ Eq.~(\ref{I_sum}) is exact if the inverse LIT of the weight function $g(\omega)$ exists, and 
$(ii)$ $I_M$ converges to $I$ at the same rate as the LIT does.
In Sec.~\ref{Existence}, 
we obtain the constraints required for the existence of the inverse LIT of $g(\omega)$, 
which follow from the properties of the response function and the analytic form of $g$.
In Sec.~\ref{Lanczos} 
we illustrate the technical details behind the derivation of Eq.~\eqref{SR_lanc}, based on the Lanczos algorithm. 
In Sec.~\ref{Results}
we show the practical convergence rate of 
the LSR method with respect to the Lanczos step $N$ for sum rules associated with the electric dipole transition,
emphasizing that a rapid convergence in $N$ can be reached for a variety of weight functions.

\section{Derivation}\label{Derivation}

As detailed in Refs.~\cite{EFROS94,REPORT07}, 
the response function can be obtained from a numerical calculation using its
Lorentz integral transform (LIT)
\begin{equation} \label{l_sgm}
{\cal L}(\sigma,\Gamma)=\frac{\Gamma}{\pi}\int d\omega
  \frac{S(\omega)}{(\omega-\sigma)^2+\Gamma^2}\;, 
\end{equation}
which is an integral transform of the response function with a Lorentzian kernel.
In order to prove the validity of evaluating the SR of Eq.~\eqref{eq:SR}
using the square-integrable basis represented by Eq.~\eqref{I_sum},
we assume that there exists a function $h(\sigma,\Gamma)$
such that the weight function $g(\omega)$ in Eq.~\eqref{eq:SR} can be presented in the form
\begin{equation}
\label{h_sgm}
  g(\omega)=\frac{\Gamma}{\pi}\int d\sigma
  \frac{h(\sigma,\Gamma)}{(\omega-\sigma)^2+\Gamma^2}\;.  
\end{equation}
Comparing Eqs.~\eqref{l_sgm} and \eqref{h_sgm} we find that the relation
between $g(\omega)$ and $h(\sigma,\Gamma)$ is similar to the relation between
${\mathcal L(\sigma, \Gamma)}$ and $S(\omega)$. 
There is, however, one important difference: 
for any physical response function, the LIT integral ${\cal L}(\sigma,\Gamma)$ is well defined. 
 In contrast, the existence of $h(\sigma,\Gamma)$ is not self evident, 
 and the conditions under which $h(\sigma,\Gamma)$ exists
will be discussed in the next section. At this moment we shall carry out our
arguments assuming that (\ref{h_sgm}) holds.

Inserting the LIT representation of the weight function (\ref{h_sgm}) into
the SR of Eq.~(\ref{eq:SR}) and changing the order of  
integration, we rewrite the SR in terms of the LIT function 
${\cal L}(\sigma,\Gamma)$ instead of $S(\omega)$
\begin{eqnarray}\label{eq:I_Lh}
  I &=& \int d\omega\, \int d\sigma \, 
        S\left(\omega\right)\frac{\Gamma}{\pi}\frac{h(\sigma,\Gamma)}{(\omega-\sigma)^2+\Gamma^2}     
\cr &=& \int d\sigma \, {\cal L}(\sigma,\Gamma) h(\sigma,\Gamma)\;.
\end{eqnarray}
Similar to Eq.~\eqref{eq:SR}, also here 
the integration limits can be regarded as effectively finite, 
due to the properties of $\cal L$.

The advantage of introducing the LIT function ${\cal L}(\sigma,\Gamma)$ stems 
from the fact that it can be
calculated numerically using a  set of localized basis functions with  bound-state-like 
boundary conditions~\cite{EFROS94,REPORT07}. Consequently, it is much simpler to calculate 
${\cal L}(\sigma,\Gamma)$ than the response function itself.
We expand ${\cal L}(\sigma,\Gamma)$ over the set of square-integrable basis functions 
$\{|\mu\ket\}$ that diagonalizes the Hamiltonian, obtaining 
\begin{eqnarray}
\label{L_epsmu_diag}
{\cal L}_M(\sigma,\Gamma)
                      & =&\frac{\Gamma}{\pi}\sum_{\mu\ne i}^M \frac{ |\bra \mu | \hat O | i\ket|^2}
                     {(\omega_{\mu}-\sigma)^2+\Gamma^2}\;.
\end{eqnarray}
Substituting the calculated ${\cal L}_M(\sigma,\Gamma)$ of Eq.~\eqref{L_epsmu_diag} into \eqref{eq:I_Lh} we recover Eq.~\eqref{I_sum} for $I_M$, 
which is to some extent an intuitive discrete representation of the SR. 
Nevertheless, with this derivation we have further justified the use of a localized basis.

If we consider an expansion on a basis of size $M$, such that the accuracy of the
calculated function ${\cal L}_M(\sigma,\Gamma)$
is within $\varepsilon_M$,
\begin{equation} 
    |{\cal L}(\sigma,\Gamma)-{\cal L}_M(\sigma,\Gamma)| \leq 
    \varepsilon_M\;,
\end{equation} 
then the accuracy of  $I_M$ calculated 
using the same basis
can be bounded by
\begin{eqnarray}\label{convcond}
  |I-I_M| 
          &\leq& \int d\sigma \, 
                 \left|{\cal L}(\sigma,\Gamma) - {\cal L}_M(\sigma,\Gamma)\right| 
                 |h(\sigma,\Gamma)|    
\cr       &\leq& 
                 \varepsilon_M 
                 \int d\sigma \,
                 |h(\sigma,\Gamma)|   \;.
\end{eqnarray} 
Therefore, if the function $h(\sigma,\Gamma)$ exists and 
the integral $\int \! d\sigma|h(\sigma,\Gamma)|$ on
the right-hand-side of Eq.~\eqref{convcond} is finite, then 
the discretized SR in Eq.~\eqref{I_sum} 
converges to the exact sum rule $I$ at the same rate as ${\cal L}_M(\sigma,\Gamma)$ converges to ${\cal L}(\sigma,\Gamma)$.
In other words, the discrete representation becomes exact when the LIT function converges to its exact value 
without any need to recover the continuum limit. 
In the following we will present a more rigorous discussion on 
the conditions for which $h(\sigma,\Gamma)$ exists and on the convergence properties of $I_M$
constrained by Eq.~\eqref{convcond}.

\section{Existence and Convergence }
\label{Existence}

In this section we will explore the conditions 
for which we can locate such a function $h(\sigma,\Gamma)$, 
which satisfies that $\int \! d\sigma|h(\sigma,\Gamma)|$ is finite, 
and the implications on the convergence of the calculated SR.

To this end, we use the Fourier transform and utilize the
Fourier representation of the  
Lorentzian kernel~\cite{Efros99}
\begin{equation}
 \frac{1}{x^2+\alpha^2}=\frac{1}{2\alpha}\int dk e^{-\alpha|k|+ikx} \;,
\end{equation}
so that we can rewrite Eq.~(\ref{h_sgm}) in the following form
\begin{equation}
\label{eq_to_trans}
  g(\omega) = \frac{1}{2\pi}\int d\sigma\, dk\, h(\sigma,\Gamma) 
            e^{-\Gamma|k|+ik(\sigma-\omega)}\;.
\end{equation}
We define the Fourier transforms of $g$ and $h$, respectively, as
\begin{eqnarray}
 \tilde{g}(k)&=&\int d\omega \;g(\omega)e^{ik\omega}\;, \cr
 \tilde{h}(k,\Gamma)&=& \int d\sigma\; h(\sigma,\Gamma) e^{ik\sigma}\;,
\end{eqnarray}
and 
obtain from Eq.~\eqref{eq_to_trans} a simple relation between them 
\begin{equation}
\label{eq:h_tilde}
\tilde{h}(k,\Gamma)=e^{\Gamma|k|}\tilde{g}(k)\;.
\end{equation}
Performing the inverse Fourier transform on Eq.~\eqref{eq:h_tilde}
we obtain the function $h(\sigma,\Gamma)$ as~\cite{Efros99}
\begin{equation} \label{h_ft}
 h(\sigma,\Gamma)=\frac{1}{2\pi}\int dk\; e^{\Gamma|k|}\tilde{g}(k) e^{-ik\sigma}\;.
\end{equation}
This relation provides us with a necessary and sufficient condition for the existence of $h(\sigma,\Gamma)$: 
If the integral on the right hand side of Eq.~(\ref{h_ft}) converges,
 $h(\sigma,\Gamma)$ is well defined; Otherwise we cannot find $h(\sigma,\Gamma)$
that fulfills Eq.~(\ref{h_sgm}) exactly.

We first provide an extreme example, where $g(\omega)$ is a Dirac delta function
$g(\omega) = \delta(\omega-\omega_0)$, 
whose Fourier transform is $\tilde{g}(k) = \exp(ik\omega_0)$. 
It is straightforward to obtain that the integral in Eq.~\eqref{h_ft} diverges in this case.

However, in many cases
the weight function $g(\omega)$ 
will be a continuous function, which can be expanded using
a complete set of square-integrable basis functions. 
In practice, in bound nuclear systems
the response function $S(\omega)$ provides a lower bound for the integral in Eq.~\eqref{eq:SR} 
at the lowest excitation energy $\omega_{th}\! \geq\! 0$, 
and vanishes for \mbox{$\omega\rightarrow\infty$}. 
Therefore, the weight function $g(\omega)$ 
only needs to be known over a finite range of $\omega$.
Consequently, the weight function $g(\omega)$ 
may be approximated by a finite number of basis functions, i.e., 
$g(\omega) \cong \sum_{1}^k c_i g_i(\omega)$,
as long as the M{\"u}ntz-Sz{\'a}sz condition~\cite{MStheorem} is fulfilled.

Here, we provide two common choices of complete basis sets, Lorentzians and Gaussians
integral bases, respectively.
In the Lorentzian basis,
represented by 
\begin{align}\label{eq:Lorentz}
g_i(\omega) = \frac{\beta_i}{\pi} \frac{1}{(\omega-\omega_i)^2+\beta_i^2}\;,
\end{align}
the Fourier transform of 
the basis function $g_i(\omega)$ yields 
\mbox{$\tilde{g}_i(k)=\exp\left(-\beta_i |k| +i k\omega_i\right)$}. 
By substituting it in Eq.~\eqref{h_ft}, we obtain the inverse transform
\begin{equation}
\label{h_i_L}
 h_i(\sigma,\Gamma)=\frac{\beta_i-\Gamma}{\pi} \frac{1}{(\sigma-\omega_i)^2+(\beta_i-\Gamma)^2}\;,
\end{equation}
only for $\beta_i \!>\! \Gamma \!>\!0$. 
Under this condition, \mbox{$\int \!d\sigma |h_i(\sigma,\Gamma)| = 1$} is finite and thus fulfills the requirement from Eq.~\eqref{convcond}. 
However, when the weight function is narrowly peaked 
at a particular value of $\omega$, 
then in order for it to be adequately described,
the expansion must include a Lorentzian function $g_i$ that has a similar narrow width, 
 i.e., a small value of $\beta_i$. 
In this case, the condition $\beta_i > \Gamma$ can become prohibitively difficult to 
satisfy, due to 
the numerical properties
of the LIT method:
the convergence rate of ${\cal L}_M(\sigma,\Gamma)$ to ${\cal L}(\sigma,\Gamma)$ 
worsens as $\Gamma \rightarrow 0$.

\begin{figure}[htb]
\centering
\includegraphics[width=0.45\textwidth,clip=true]{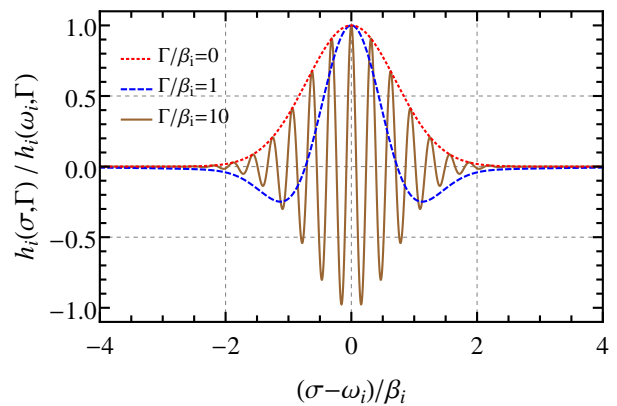}
\caption{(Color online) The Gaussian-basis-related function $h_i(\sigma,\Gamma)$ of Eq.~\eqref{h_i_G} is plotted as a function of the argument 
$(\sigma-\omega_i)/\beta_i$ for various values of $\Gamma/\beta_i$. 
$h_i(\sigma,\Gamma)$ is rescaled by division with its peak value at $\sigma=\omega_i$.}
\label{fig:h_i_Erf}
\end{figure}

For the Gaussian basis,
expressed as 
\begin{align}\label{eq:Gauss}
g_i(\omega) = \frac{1}{\sqrt{\pi}\beta_i} e^{-(\omega-\omega_i)^2/\beta_i^2}\;,
\end{align}
the Fourier transform of 
the basis function $g_i(\omega)$
leads to $\tilde{g}_i(k) = \exp(-\frac{\beta_i^2 k^2}{4} + i k \omega_i)$. 
Similarly, we substitute it into Eq.~\eqref{h_ft}, and find the 
inverse transform in this case to be
\begin{equation}
\label{h_i_G}
h_i(\sigma,\Gamma)=\frac{1}{\sqrt{\pi}\beta_i}\,
\mathrm{Re}\left[ \mathcal{F}(z) \right] \; ,
\end{equation}
with
\begin{equation}
\mathcal{F}(z)\equiv \left[1+ \rm{Erf}\left( z \right)\right]\exp{(z^2)}
\end{equation}
and
$z\equiv \frac{\Gamma+i\left(\sigma-\omega_i\right)}{\beta_i}$.
The function $\rm{Erf}(z)$ is the error function defined by
\begin{equation}
\rm{Erf}(z) = \frac{2}{\sqrt{\pi}} \int\limits^z_0 e^{-t^2} d t \;.
\end{equation}
As shown numerically in Fig.~\ref{fig:h_i_Erf}, $h_i(\sigma,\Gamma)$
is symmetric around its peak value at $\sigma=\omega_i$, i.e., 
$|h_i(\sigma,\Gamma)| \leq h_i(\omega_i,\Gamma) = \mathcal{F}(\Gamma/\beta_i) / \sqrt{\pi {\beta_i}^2}$,
and vanishes rapidly as the value of $|\sigma-\omega_i|$ increases.
Therefore, one can demonstrate that the integral $\int \!d\sigma |h_i(\omega_i,\Gamma)|$ will be finite, 
provided that $h_i$'s peak value, $h_i(\omega_i,\Gamma)$, is finite. 
However, $h_i(\omega_i,\Gamma)$ grows extremely fast with increasing $\Gamma/\beta_i$, 
because of its $\exp(\Gamma^2/\beta_i^2)$ component, and so does the integral $\int \! d\sigma |h_i(\omega_i,\Gamma)|$.
Consequently, if $g(\omega)$ can be adequately described by a set of Gaussian basis functions, then 
the corresponding function $h(\sigma,\Gamma)=\sum_1^k c_i h_i(\sigma,\Gamma)$ can be constructed, 
and the convergence condition 
required by Eq.~\eqref{convcond} holds if $\Gamma/\beta_i$ is kept small.
However, again it is evident that calculation of SRs associated
with narrowly peaked weight functions may become prohibitively difficult. 
This situation, manifested in the Gaussian basis by the presence of a $g_i(\omega)$ with small $\beta_i$, 
makes it challenging to find an appropriate $\Gamma$ that is large enough to make ${\cal L}_M$ converge at a reasonable rate, 
and small enough to keep $\exp(\Gamma^2/\beta_i^2)$ small.

These examples indicate the robustness of the LSR method in the general case. 
However, the short coming of the method is to resolve SRs 
whose energy weight contain very narrow features or discontinuities.
In practice, the convergence of the results need to be 
verified numerically for each case individually.

\section{The Lanczos algorithm}
\label{Lanczos}

The calculation of the SR 
in Eq.~(\ref{I_sum}),
as well as the calculation of the LIT of Eq.~(\ref{L_epsmu_diag}), involves a
summation over the entire spectrum of ${\mathcal H}$  using a discretized complete basis expansion.
However, the basis size $M$ required to obtain converged results is usually 
too large to allow a direct diagonalization of ${\mathcal H}$.
To overcome this problem in the LIT method, 
the use of the Lanczos algorithm was introduced~\cite{Mario}. 
This practice has been successfully applied to many electroweak processes of light nuclei~\cite{REPORT07}.
In these cases, it was found that with sufficiently large number of Lanczos steps, 
which is still much smaller than the basis size, 
${\cal L}_M$ can converge to the exact value ${\cal L}$  within a sub-percentage accuracy.

Following this example, we seek an equivalent formalism for calculating the SR using
the eigenstates of the Hamiltonian in
the Krilov subspace~\cite{Saad2003}
 $\{|\varphi_0\ket, {\mathcal H} |\varphi_0\ket, {\mathcal H}^2 |\varphi_0\ket,
\ldots {\mathcal H}^{N-1} |\varphi_0\ket \}$. 
The appropriate choice of the starting vector 
$|\varphi_0\ket$ is the normalized transition vector
\begin{equation} |\varphi_0\ket  = \hat O |i\ket / 
    \sqrt{\bra i | \hat O^\dagger \hat O |i\ket} \;.
\end{equation}
Using the Lanczos algorithm \cite{Lanczos} one can straightforwardly 
construct an orthonormal basis 
$\{|\varphi_0\ket, |\varphi_1\ket, |\varphi_2\ket,\ldots |\varphi_{N-1}\ket \}$ 
to the Krilov subspace. 
In the Lanczos algorithm one represents $\mathcal{H}$, 
which is an $M\times M$ matrix, 
by the tridiagonal $N\times N$ matrix
\begin{equation}\label{eq: 5-H_N}
   \hat{H}_{N}=\left(\begin{array}{ccccc}
   a_{0} & b_{1} & 0 & \cdots & 0\\
   b_{1} & a_{1} & b_{2} & \ddots\\
   0 & b_{2} & a_{2} & \ddots\\
   \vdots & \ddots & \ddots & \ddots & b_{N-1}\\
   0 &  &  & b_{N-1} & a_{N-1}
   \end{array}\right)\;,
\end{equation}
where $N\leq M$. 
$\hat{H}_{N}$ and $\{|\varphi_i\ket\}$ are constructed 
through the Lanczos iterations,
\begin{equation} 
   b_{i+1}|\varphi_{i+1}\rangle =\mathcal{H}|\varphi_i\rangle - 
   a_i|\varphi_i\rangle -b_i|\varphi_{i-1}\rangle \;,
\end{equation}
where the coefficients $a_i$ and $b_i$ are obtained as
\begin{equation}
a_i = \bra \varphi_i | \mathcal{H} |\varphi_i\ket~; \;\;\; b_i = \| b_i |\varphi_i\ket \|\;.
\end{equation}
This representation effectively provides 
the distribution of the starting vector $|\varphi_0\ket$ on the 
low-lying spectrum of $\mathcal{H}$~\cite{Haxton}.

The tridiagonal matrix $\hat{H}_{N}$ can be easily diagonalized through unitary transformation
\begin{equation}
\hat{H}_{N}=Q D Q^{\dagger}\;,
\end{equation}
where $D \equiv diag\left( E^{(N)}_0,E^{(N)}_1, \ldots ,E^{(N)}_{N-1}\right)$.
The first $N$ eigenstates of $\mathcal H$ approximated by the Lanczos algorithm 
are therefore given by 
\begin{equation}
|\tilde{\nu}\ket = \sum\limits_{j=0}^{N-1} Q_{j\nu}\, |\varphi_j\ket \;,
\end{equation}
where $|\tilde{\nu}\ket$ satisfies 
\mbox{$\mathcal{H} |\tilde{\nu}\ket \cong \hat{H}_{N} |\tilde{\nu}\ket = E_{\nu}^{(N)} |\tilde{\nu}\ket$}.
If we now substitute the discrete sum over the states $|\mu\ket$ in Eq.~\eqref{I_sum} 
with a sum over the $N$ approximated eigenstates $|\tilde{\nu}\ket$, we obtain
\begin{eqnarray}
\label{SR_lanc_2}
I_N &\equiv & \sum_{\nu=0}^{N-1}|\bra \tilde{\nu}| \hat O | i\ket|^2
    g(\omega_\nu) \cr
    &= & \bra i |\hat O^\dagger \hat O |i\ket
         \sum_{\nu=0}^{N-1} \,
         \left| \sum_{j=0}^{N-1}  Q_{j\nu}^\dagger \bra \varphi_j | \varphi_0\ket \right|^2
    g(\omega_\nu) \cr
    &= & \bra i |\hat O^\dagger \hat O |i\ket 
         \sum_{\nu=0}^{N-1}| Q_{\nu 0}|^2 
    g(\omega_\nu) \;,
\end{eqnarray}
which yields Eq.~\eqref{SR_lanc} given in the Introduction.
In this form, the calculated SR
depends only on the norm of the transition vector
and on the eigenstates and eigenvalues of the Lanczos matrix $\hat{H}_{N}$.
Furthermore, this sum converges very rapidly with increasing $N$, 
due to the excellent convergence properties of the Lanczos algorithm~\mbox{\cite{Kaniel1966,Paige1971,Saad1980}}.
In practice we have found that a few hundred Lanczos steps $N$, which is much less than the basis size $M$, 
are sufficient for convergence of the most commonly used SRs.

We would like to add two comments regarding the application of 
Eq.~\eqref{SR_lanc_2}: 
(i) If one is only interested in the inelastic response,
then the initial state should be excluded from the sum in Eq.~\eqref{SR_lanc_2}. 
(ii) The SR defined in Eq.~\eqref{SR_lanc_2} is equivalent to the ground state expectation value
\mbox{$\bra i |\hat O^\dagger g\left( \hat{H}_{N} - E_i\right) \hat O |i\ket$}, 
which is finite even with functions $g(\omega)$ 
for which the integral~\eqref{eq:SR} does not exist.

\section{Results}
\label{Results}
As an example for the application of the proposed LSR method, 
we consider the unretarded dipole response function of the $^4$He nucleus, 
calculated using the effective interaction hyperspherical harmonics (EIHH) method~\cite{EIHH,EIHH_3NF}.
This response function was first calculated for $^4$He 
using a modern nuclear Hamiltonian
in Ref.~\cite{Doron2006}.
More recently, it was used in Ref.~\cite{Ji2013a} to estimate nuclear polarizability effects in muonic~$^4$He.
Here we use the same model space as in~\cite{Ji2013a}.
 In the Hamiltonian ${\mathcal H}$ we adopt the chiral effective field theory potential, 
 including two-nucleon interactions up to next-to-next-to-next-to-leading order, 
 and three-nucleon interactions up to next-to-next-to-leading order~\cite{Machleidt:2011zz}, 
 with parametrization from Ref.~\cite{Na07}.
The dipole response function $S_{D_1}$ is defined as in Eq.~(\ref{eq:S_omg}) 
where the operator $\hat O$ is
the unretarded dipole operator
\begin{equation}
\hat D_1 \equiv \frac{1}{Z}\sum_i^Z R_i Y_1(\hat R_i) \;.
\end{equation}
Here, ($R_i$, $\hat R_i$) denotes the position of the $i$-th proton in the center-of-mass frame, 
and $Y_1$ denotes the rank-one spherical harmonics.

\begin{figure}[htb]
\centering
\includegraphics[scale=0.8,clip=]{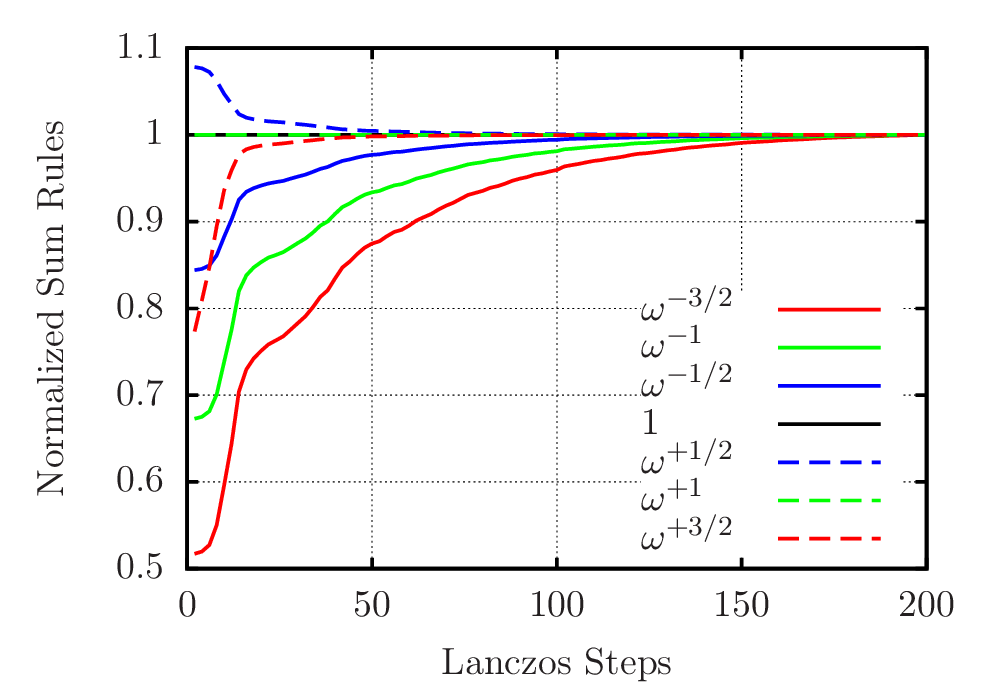}
\caption{(Color online) Convergence of various energy-dependent
sum rules with increasing number of Lanczos steps. 
The calculations are done using the unretarded dipole response of $^4$He. 
The sum rules are normalized to their converged values.} 
\label{fig:conv}
\end{figure}

In Fig.~\ref{fig:conv} we demonstrate the convergence of the LSR method for dipole sum rules
with weight functions
\mbox{$g(\omega)=\omega^{-3/2},\omega^{-1},\ldots ,\omega^{3/2}$} 
defined as
\begin{equation}
\mathcal{I}_{D_1}^{(n)} = \int_{\omega_{th}}^\infty d\omega\, \omega^{n}\, S_{D_1}(\omega) \;. 
\label{ESR}
\end{equation}
It can be seen that when calculating these SRs as in Eq.~(\ref{SR_lanc_2}), 
$N\approx200$ Lanczos steps are sufficient for convergence.
This should be compared with $M\approx 10^5$,
the size of the model space used in this calculation. 
Indeed, a slightly slower convergence is obtained 
for more negative powers of $\omega$, since $g(\omega)$ becomes relatively narrower at the origin. 
However, this does not create practical difficulties  to obtain convergence.
These examples demonstrate the accuracy and efficiency of the LSR method.

We would like to add that the dipole SR with \mbox{$g(\omega)=\omega^{-1}$} 
is related to the electric dipole polarizability
$\alpha_E$ through $\alpha_E = (8\pi Z^2 \alpha /9) \mathcal{I}_{D_1}^{(-1)}$. 
Our result for $\alpha_E$ of $^4$He is $0.0694$ fm$^{3}$, 
which is in agreement with a previous calculation~\cite{Stetcu:2009py}, 
considering the accuracy estimated there.

\begin{center}
\begin{table}[ht!]
\caption{Energy-dependent SRs of the unretarded dipole response of $^4$He.
A comparison between the Lanczos sum-rule method (LSR) and
  an explicit integration of the response function (ESR). 
  Both calculations are made with the same model space as in 
  ~\cite{Ji2013a}. 
  }
\label{tbl:compare}
\begin{tabular}{llccc}
\hline\hline
Weight & Units & LSR & ESR & Relative Difference \\ 
\hline
$\omega^{-3/2}$ & fm$^{7/2}$     & 8.414          & 8.434      & $\sim 10^{-3}$\\ 
$\omega^{-1}$   & fm$^3$ $\,$    & 3.405          & 3.409      & $\sim 10^{-3}$\\ 
$\omega^{-1/2}$ & fm$^{5/2}$     & 1.431          & 1.431      & $\sim 10^{-4}$\\ 
${1}$           & fm$^2$         & 0.637          & 0.636      & $\sim 10^{-3}$\\ 
$\omega^{1/2}$  & fm$^{3/2}$     & 0.312          & 0.309      & $\sim 10^{-2}$\\ 
$\omega$        & fm$^{1}$       & 0.178          & 0.173      & $\sim 10^{-2}$\\ 
$\omega^{3/2}$  & fm$^{1/2}$     & 0.133          & 0.117      & $\sim 10^{-1}$\\
\hline\hline
\end{tabular}
\end{table}
\end{center}

In Table~\ref{tbl:compare}
we compare the energy-dependent SRs calculated using the LSR method with SRs obtained from 
an explicit integration of the response function (ESR) as in Eq.~(\ref{ESR}),
for various weight functions $g(\omega)$. 
It can be seen that there is an excellent agreement between the two methods.
One should keep in mind that when the  many-body Hamiltonian needs to be solved, as in this example, 
the calculation of the response function is always limited in accuracy. 
Therefore, it is remarkable that the relative difference between the results of 
the two methods is rather small for most SRs. 
In this example, the response function used for explicit integration
was only calculated up to $\omega_{max}=518$~MeV.
This leads to a varying error in the ESR results.
Obviously, this error is larger for weight functions with higher powers of $\omega$, 
as can be seen from the difference between the ESR and 
LSR results for $g(\omega)=\omega^{3/2}$ in Table~\ref{tbl:compare}.
We note that the numerical accuracy of the LIT method,
and therefore also of the LSR,
is at the sub-percentage level (see Ref.~\cite{Ji2013a}).

\section{Summary}
We have presented an efficient algorithm for evaluating a 
generalized energy-dependent sum rule and discussed its properties and applicability. 
With the proposed method one can calculate sum rules without explicitly calculating the continuum states 
or the response function, thus achieving better accuracy.
We have shown that the convergence rate 
of the calculated sum rule is 
bounded by the convergence rate of the LIT function 
times a factor, 
whose magnitude
depends on the structure of the weight function.
A set of localized basis functions can be used to calculate the sum rule, 
and using the Lanczos algorithm we have demonstrated that a few hundred steps of Lanczos iterations are
sufficient to achieve a reasonable accuracy. 
In fact, this is an order of magnitude lower than the number of Lanczos steps needed 
to calculate the corresponding response function using the LIT method.
This suggests that the proposed method can be valuable also in other cases,
for example to calculate polarization corrections in other muonic atoms, as well as 
any energy-dependent sum rule 
that is of relevance in electro-weak reactions of nuclei.

\begin{acknowledgments}
  We are indebted to Giuseppina Orlandini and Victor D. Efros for useful conversations
  and valuable comments on the manuscript.
  This work was supported in parts by the Natural Sciences
  and Engineering Research Council (NSERC), the National Research
  Council of Canada, the Israel Science Foundation (Grant number
  954/09), and the Pazi foundation. 
\end{acknowledgments}

\end{document}